\documentclass[cits]{PoS}

\def\fm {\mathop{\hbox{fm}}}
\def\MeV {\mathop{\hbox{MeV}}}

\usepackage{amsmath} 
\usepackage{dsfont}  
\usepackage{bm}      

\def\beq{\begin{equation}}
\def\eeq{\end{equation}}
\def\beqs#1\eeqs{\beq\begin{split} #1 \end{split}\eeq}

\def\comment#1{}

\def\norm#1{\left\| #1 \right\|}

\def\opbraket#1#2#3{ \left\langle #1 \left| #2 \right| #3 \right\rangle}

\graphicspath{{figures/}}

\title{Chiral Polarization Scale at Finite Temperature}

\ShortTitle{Chiral Polarization Scale at Finite Temperature}

\author{\speaker{Andrei Alexandru}\\
       The George Washington University, Washington, DC, USA\\
       E-mail: \email{aalexan@gwu.edu}}

\author{Ivan Horv\'ath\\
        University of Kentucky, Lexington, KY, USA\\
        E-mail: \email{horvath@pa.uky.edu}}

\abstract{We study the chiral polarization properties of low-lying Dirac eigenmodes at 
finite temperature using the overlap operator. Results for pure gauge theory on both sides of 
deconfinement phase transition are presented. We find that the polarization scale decreases 
as we increase the temperature, but it remains non-zero as we cross in the deconfined phase 
and vanishes only when $T\approx 1.4 T_c$. This is caused by the presence of 
near-zero modes which, we find, are chirally polarized.}

\FullConference{The 30th International Symposium on Lattice Field Theory\\
		 June 24 -- 29,  2012\\
		 Cairns, Australia}

\begin{document}

\section{Introduction}

Banks-Casher relation connects the low-lying spectrum of the Dirac operator to the spontaneous
symmetry breaking in QCD~\cite{Banks:1979yr}. This relationship is rather generic and a more detailed
understanding of the mechanism responsible for the chiral symmetry breaking is thought to be
encoded in the chiral properties of the low lying eigenmodes. If we separate the chiral components
of the Dirac eigenmode $D\psi = \lambda\psi$,
\beq
\psi_R =\frac12(1+\gamma_5)\psi  \qquad\text{and}\qquad \psi_L=\frac12(1-\gamma_5)\psi\,,
\eeq
the relative magnitudes of these components at every lattice point, i.e., $\norm{\psi_{R,L}(x)}$,
carry information about the {\em local chirality} of the mode. For the eigenmodes of the
free Dirac operator we have $\opbraket{\psi}{\gamma_5}{\psi}=0$ and, using translational
symmetry for the chiral components magnitude, we can show that $\norm{\psi_R(x)}=\norm{\psi_L(x)}$.
We say that these modes are anti-polarized since their left and right components have equal magnitude
at every point.

In the presence of a gauge background, the chiral components of the eigenvector satisfy the following
equations
\beqs
\left[ -D^2+\frac12\sigma_{\mu\nu}F^S_{\mu\nu} \right]\psi_L = \lambda^2\psi_L \,,
\quad F^S = \frac12 \left( F+\tilde F \right) \,,\\
\left[ -D^2+\frac12\sigma_{\mu\nu}F^A_{\mu\nu} \right]\psi_R = \lambda^2\psi_R \,,
\quad F^A = \frac12 \left( F-\tilde F \right) \,,
\eeqs
where $\tilde{F}_{\mu\nu} = \frac12 \epsilon_{\mu\nu\alpha\beta} F_{\alpha\beta}$ is the dual
gauge field tensor. Note that the equations above are similar to Schr\"odinger equations in
four dimensions with $F^{S,A}$ playing the role of potential energy. 
For classical solutions of the gauge field equations, the gauge field itself is polarized, i.e.,
the field is either self-dual ($F^A=0$) or anti-self-dual ($F^S=0$)~\cite{Schafer:1996wv}. 
If the semi-classical 
approximation were relevant for QCD, one would expected that this tendency for polarization will not
be destroyed by quantum fluctuations and there would be regions of the field where
$F^A$ is strong and $F^S$ is weak and other regions where the situation is reversed.  
The equations above then imply that this tendency would also be reflected in the local
chirality of the eigenmodes~\cite{Horvath:2001ir}.

In a series of 
papers~\cite{Horvath:2001ir,Draper:2004id,Alexandru:2010sv,Alexandru:2010rg}, a method to measure
the effect of QCD dynamics on local chirality was proposed. Using this
{\em dynamical polarization}, we found that for low-lying
eigenmodes of the Dirac operator there is a small tendency for 
polarization, which turns into an anti-polarization tendency as we increase the magnitude
of the eigenvalues~\cite{Alexandru:2010sv,Alexandru:2010rg}. 
The scale where the polarization turns into anti-polarization is the {\em chiral polarization scale}.
We computed this scale on a series of quenched ensembles and showed that it survives in the
continuum limit~\cite{Alexandru:2010sv,Alexandru:2010rg}. 
The weak polarization of the low-lying modes is
similar to the polarization observed for the dual components of the gauge 
field~\cite{Alexandru:2011yy,Alexandru:2011tu}.

The studies mentioned above were carried out for QCD at zero temperature where it is well known
that the chiral symmetry is broken. If the polarization of the low-lying eigenmodes is
indeed related to chiral symmetry breaking, we expect it to vanish when the symmetry
is restored. In this study we compute the chiral polarization scale as we increase
the temperature going from the chirally broken phase at low temperature to the 
chirally symmetric phase at high temperatures.

The plan of the paper is the following. In Section~\ref{sec:tools} we briefly review the tools 
used in this investigation: dynamical polarization and chiral polarization scale. 
In Section~\ref{sec:results} we discuss the ensembles used in this study and show
our results. Finally, in Section~\ref{sec:conclusions} we present our conclusions.

\section{Dynamical polarization and chiral polarization scale}
\label{sec:tools}

Dynamical polarization can be defined in a general context~\cite{Alexandru:2010sv}, but we
will discuss it here in connection with local chirality. For a given set of 
eigenmodes of the Dirac operator evaluated on an ensemble of gauge configurations, each
lattice point produces a pair of chiral components $q_{1,2}=\norm{\psi_{R,L}(x)}$. The
probability distribution for these pairs is denoted with ${\cal P}(q_1, q_2)$. To measure
the polarization of a given pair, one can use {\em polarization variable} 
$X=X(q_1, q_2)$, for example 
\beq
X = \frac4\pi\arctan \frac{q_2}{q_1} - 1 \,,
\eeq
the polar angle in $(q_1, q_2)$ plane rescaled to the interval $[-1,1]$.
The probability distribution of $X$ in ${\cal P}$ assesses the degree of local chirality.
However, this approach is kinematical since the shape of the final histogram is 
determined by the choice of polarization variable $X$.

\begin{figure}[!t]
\center{
\includegraphics[width=7cm]{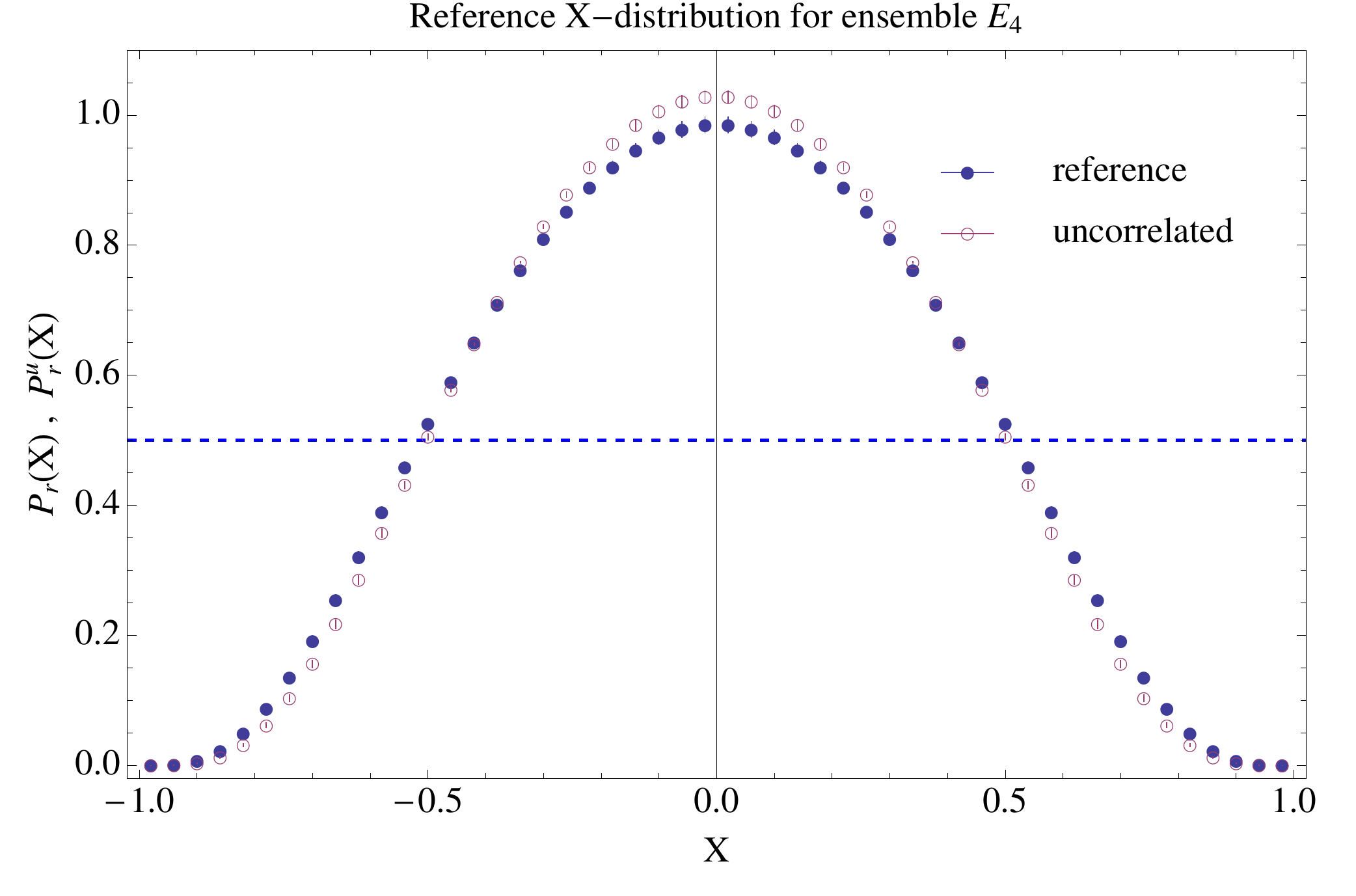}
\includegraphics[width=7cm]{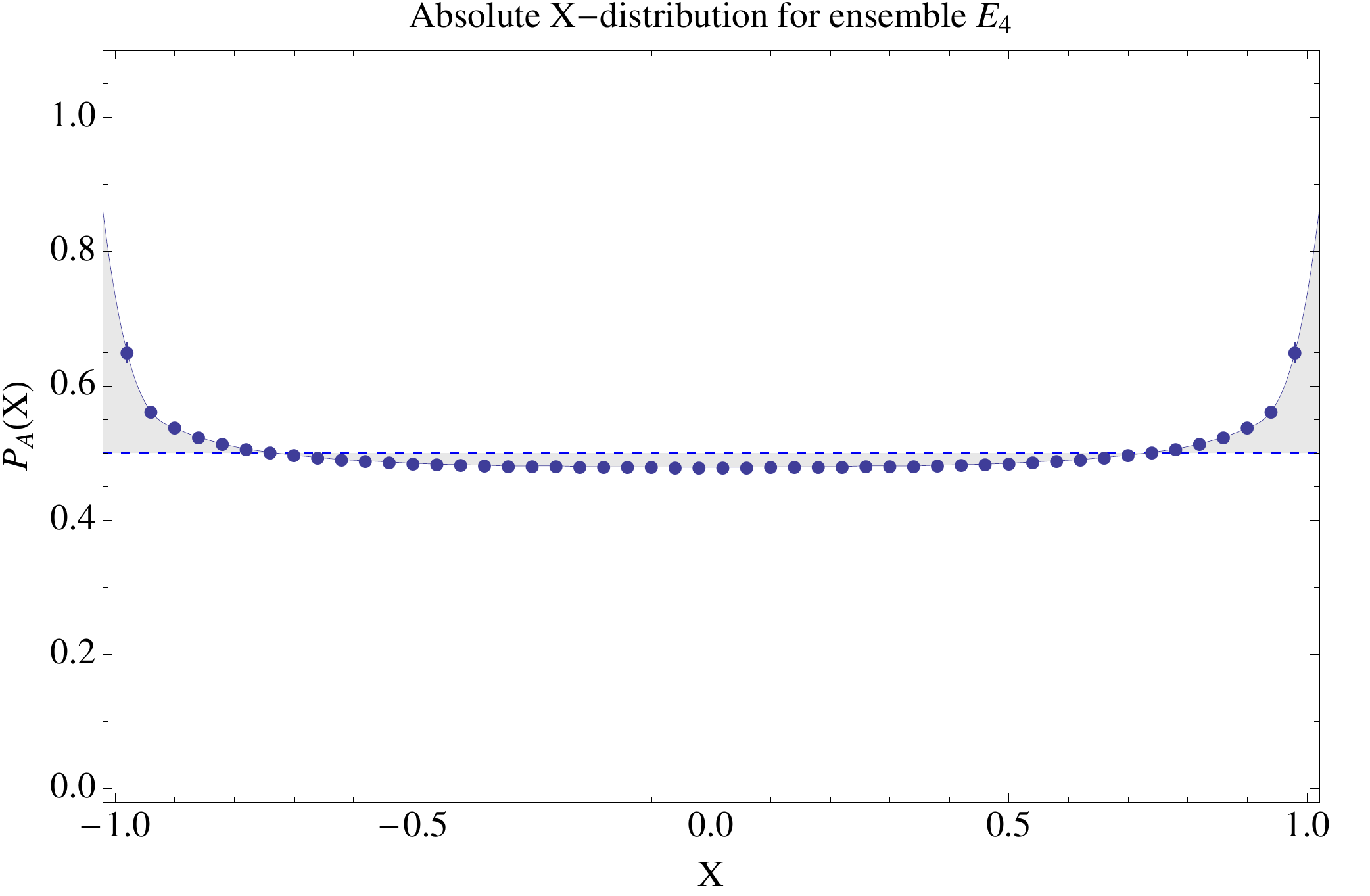}
}
\caption{Left panel: The distribution of polarization variable in ${\cal P}$ (reference)
and ${\cal P}^u$ (uncorrelated) for the lowest eigenmodes in ensemble $E_4$~\cite{Alexandru:2010sv}. 
Right panel: The absolute X-distribution 
for the same eigenmodes.}
\label{fig:dyn-pol}
\end{figure}

To gauge the polarization induced by QCD dynamics, we compare ${\cal P}$ with a similar
distribution, ${\cal P}^u$, where the correlation between the components is removed. The 
{\em uncorrelated distribution} ${\cal P}^u(q_1, q_2) = P_1(q_1)P_2(q_2)$ is generated
using the marginal distributions $P_1(q_1) = \int dq_2 {\cal P}(q_1, q_2)$
and $P_2(q_2) = \int dq_1 {\cal P}(q_1, q_2)$. Note that due to symmetries of QCD we have $P_1 = P_2$.
For example, in the left panel of Fig.~\ref{fig:dyn-pol} we plot the histogram of one
polarization variable $X$ for both correlated and uncorrelated distributions. Note that
the correlated distribution is very similar to the uncorrelated one, but it is higher
towards the extremal points and depressed in the middle. This is exactly what we expect
from a polarized distribution. To better gauge this tendency, we define the 
{\em absolute X-distribution} using the polarization variable in which the uncorrelated 
distribution is constant. This distribution will 
be peaked towards the edges for polarized case; for anti-polarized case
the absolute X-distribution will peak towards the middle. In the right panel of 
Fig.~\ref{fig:dyn-pol} we plot the absolute X-distribution for the same ensemble. We see
that this indeed confirms that the correlated distribution is polarized.

\begin{figure}[!t]
\center{
\includegraphics[width=8.3cm]{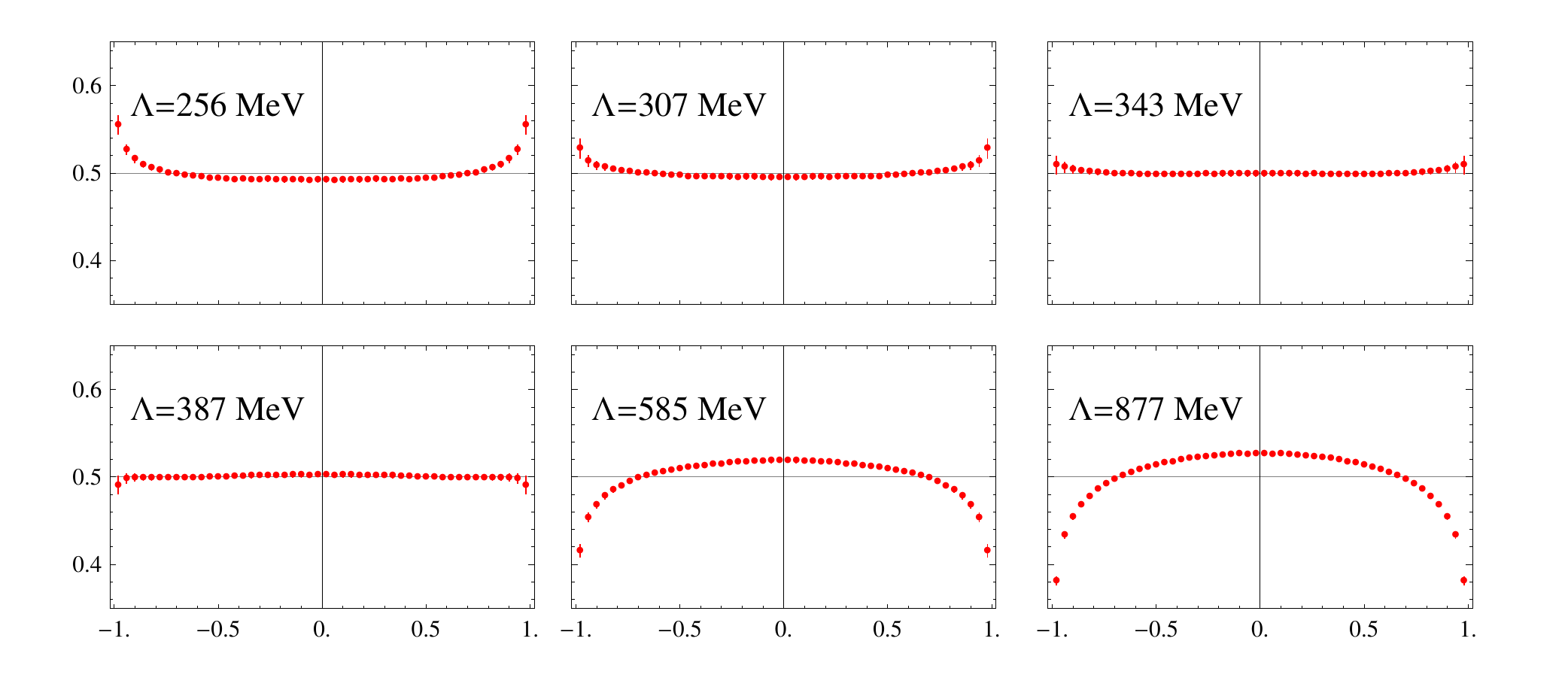}
\includegraphics[width=5.7cm]{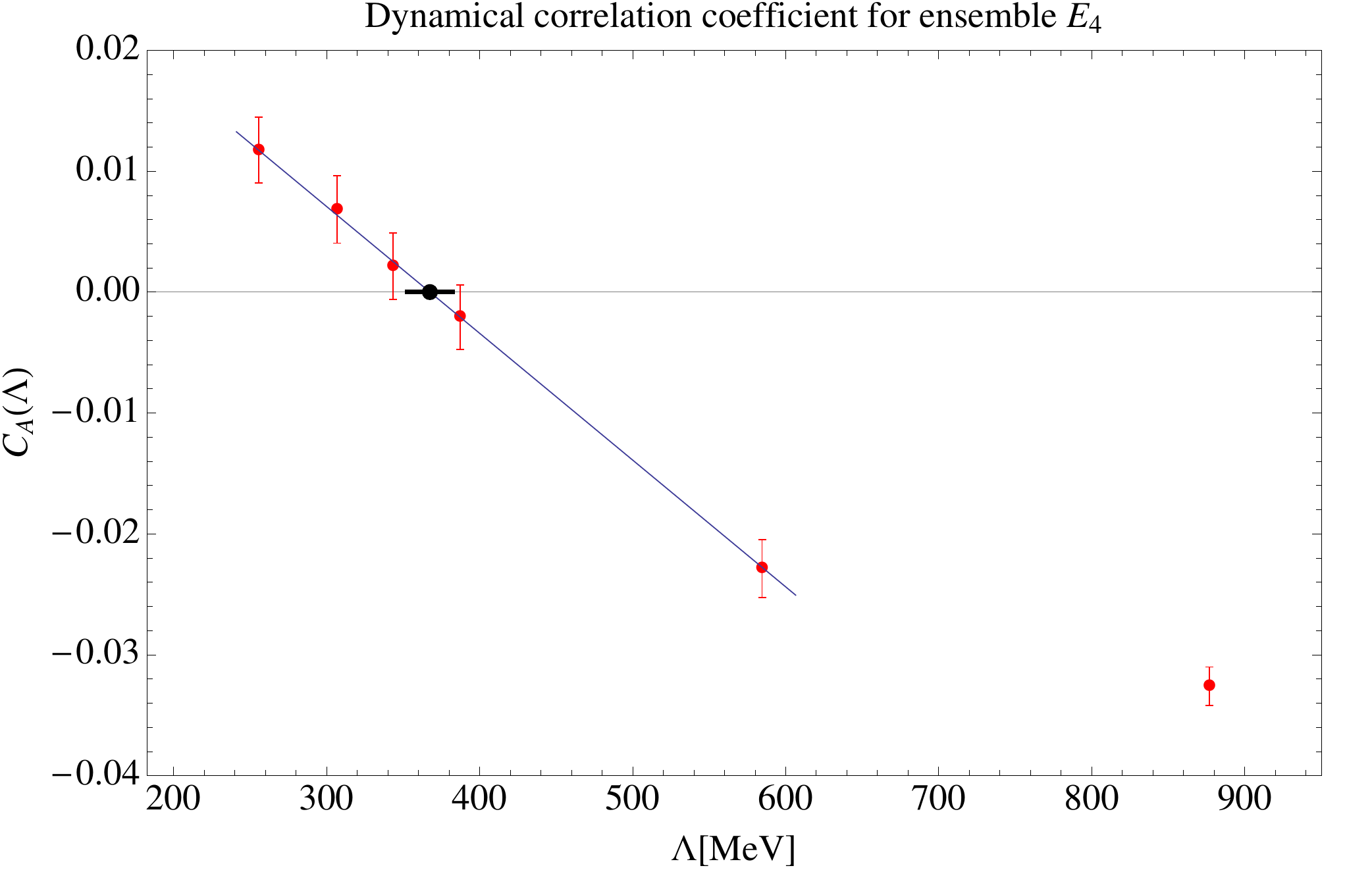}
}
\caption{Left panel: Absolute X-distribution as a function of scale for the same ensemble
as in Fig.~\protect\ref{fig:dyn-pol}.
Right panel: Correlation coefficient as a function of scale.}
\label{fig:pol-scale}
\end{figure}

As we noted in the introduction, the low-lying modes of the Dirac operator in QCD at
zero temperature are weakly polarized~\cite{Alexandru:2010sv}. In the left panel of Fig.~\ref{fig:pol-scale} 
we plot the absolute X-distribution for eigenmodes at different scales. As we increase 
the eigenvalue this tendency weakens and somewhere in the interval $343\MeV\leq\Lambda\leq387\MeV$ the 
polarization vanishes. To define the {\em chiral polarization scale} where the polarization vanishes, we use
the correlation coefficient of chiral polarization~\cite{Alexandru:2010sv}
\beq
C_A = 2\Gamma - 1 \quad\text{with}\quad \Gamma = \int_{-1}^1 dX |X| P_A(X) \,, 
\eeq
where $P_A(X)$ is the absolute X-distribution. Note that
$\Gamma$ is the probability that a pair drawn from ${\cal P}$ is more polarized than
another independently drawn from ${\cal P}^u$. If ${\cal P}$ is the same as ${\cal P}^u$,
we have no polarization and $\Gamma=1/2$. The correlation coefficient is normalized
to be $0$ in this situation, positive for polarized distributions and negative for
anti-polarized ones. In the right panel of Fig.~\ref{fig:pol-scale} we plot
the correlation coefficient, which allows us to easily extract the chiral polarization
scale for the ensemble, $\Lambda_T=368(15)\MeV$.

\section{Results}
\label{sec:results}

To explore the connection between eigenmodes' polarization and chiral symmetry breaking,
we measured the chiral polarization scale at different temperatures, both below and
above the deconfinement transition. We generated a set of quenched ensembles using
Wilson gauge action with $\beta=6.054$. This value of $\beta$ corresponds to a lattice spacing of 
$a/r_0=0.170$ according to a non-perturbative parametrization~\cite{Guagnelli:1998ud}. 
Using $r_0=0.5\fm$ we get $a=0.085\fm$. To avoid issues related to changing the cutoff, we
varied the temperature by changing the temporal extent while keeping the lattice spacing fixed.
The spatial volume was kept fixed at $V=(20 a)^3=(1.7\fm)^3$.
The parameters for our ensembles are presented in Table~\ref{table:ens}.

\begin{table}[b]
\begin{center}
\begin{tabular}{l|cccccccc}
$N_t$ & 4 & 6 & 7 & 8 & 9 & 10 & 12 & 20 \\
\hline
$N_\text{conf}$ & 100 & 100 & 400 & 400 & 200 & 200 & 200 & 100 \\
\hline
$T/T_c$ & 2.09 & 1.39 & 1.20 & 1.05 & 0.93 & 0.84 & 0.70 & 0.42
\end{tabular}
\end{center}
\caption{Parameters for the ensembles used in this study. The temperature is determined using
the lattice spacing and the critical temperature $T_c=277\MeV$~\cite{Karsch:2003jg}.}
\label{table:ens}
\end{table}%

\begin{figure}[!t]
\center{
\includegraphics[height=5.225cm]{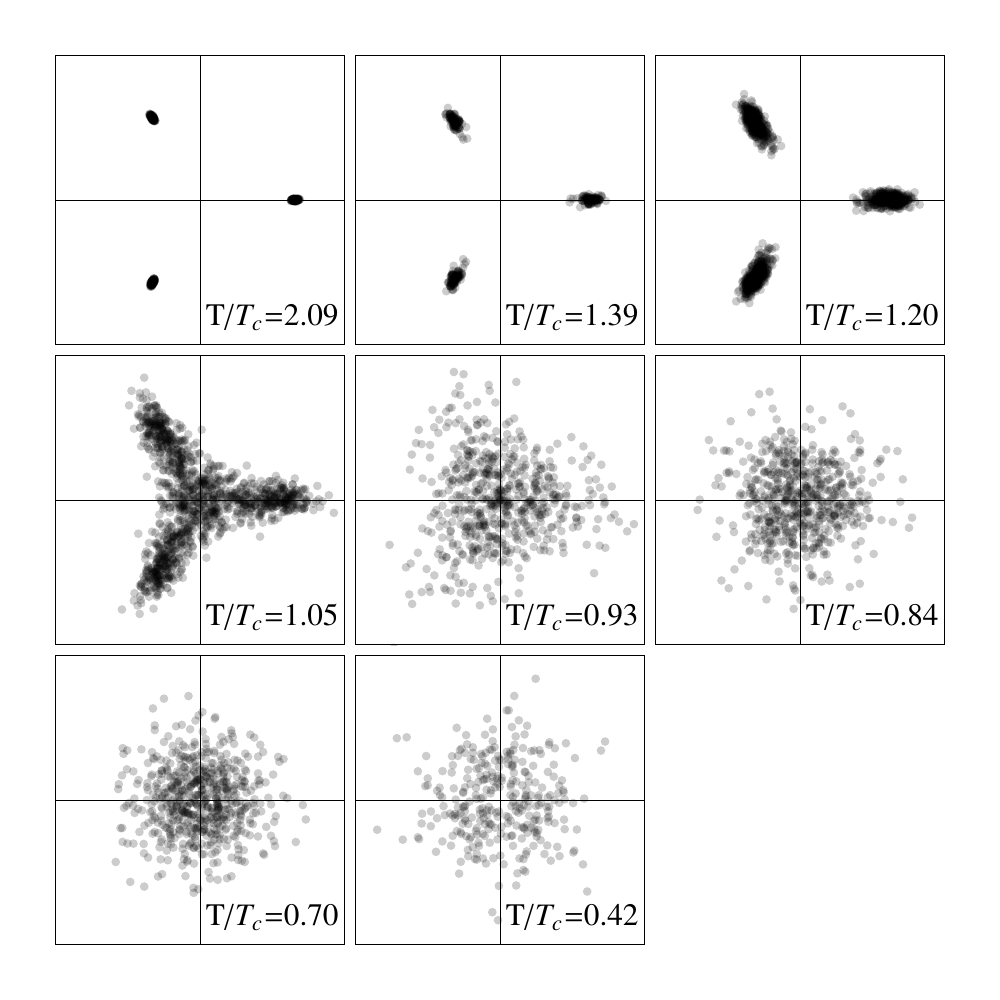}
\includegraphics[height=5.5cm]{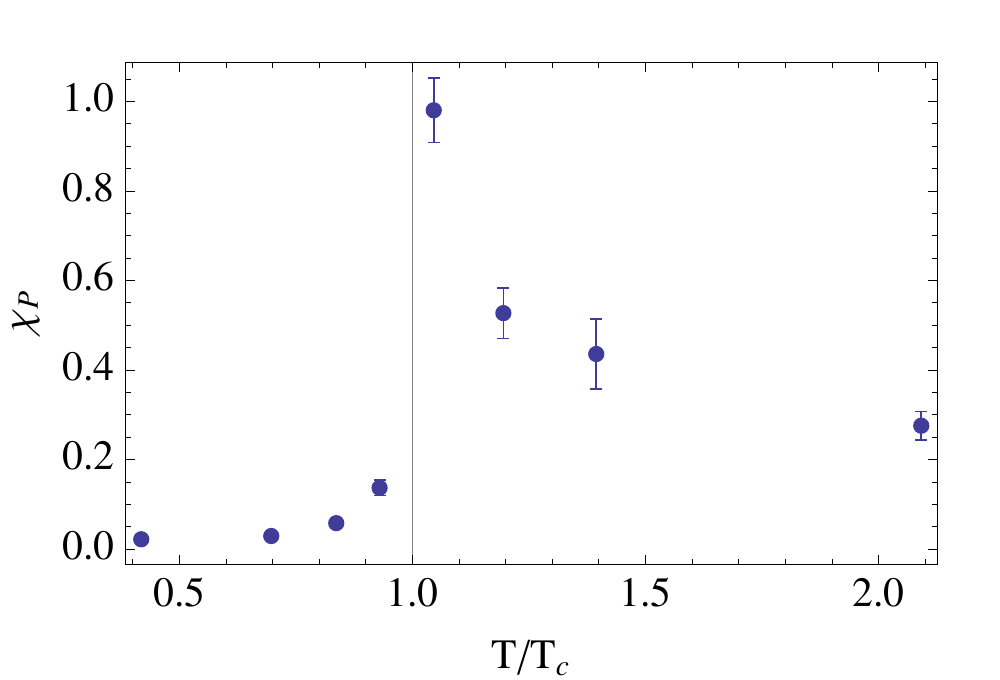}
}
\caption{Polyakov loop distribution (left) and susceptibility (right) for our ensembles.}
\label{fig:polyakov}
\end{figure}

Since there is an uncertainty in scale setting of the order of 5\%, the temperature is
determined in units of string tension~\cite{Karsch:2003jg} and the lattice spacing in units 
of $r_0$~\cite{Guagnelli:1998ud}, 
we decided to compute the Polyakov loop susceptibility to confirm that the transition 
temperature is determined correctly. In Fig.~\ref{fig:polyakov} we plot the distribution of 
the Polyakov loop on our ensembles in the left panel and the susceptibility in the right panel. 
From these figures is apparent that we have the temperature scale determined correctly. 

Pure glue theory has a $Z(3)$ symmetry that is reflected in the distribution of the 
Polyakov loop, as can be seen from the left panel of Fig.~\ref{fig:polyakov}. 
In the deconfined phase this symmetry is spontaneously broken and
the spectrum of the Dirac operator is very different for configuration from different
$Z(3)$ sectors~\cite{Chandrasekharan:1995gt}. Dynamical quarks break this 
symmetry explicitly and bias the theory towards 
configurations with Polyakov loop in the real sector ($-\pi/3 \le \arg P \le \pi/3$). 
Since for the full theory only the real sector 
is relevant, in this study we only use configurations in this sector.

\begin{figure}[!t]
\center{
\includegraphics[width=8.8cm]{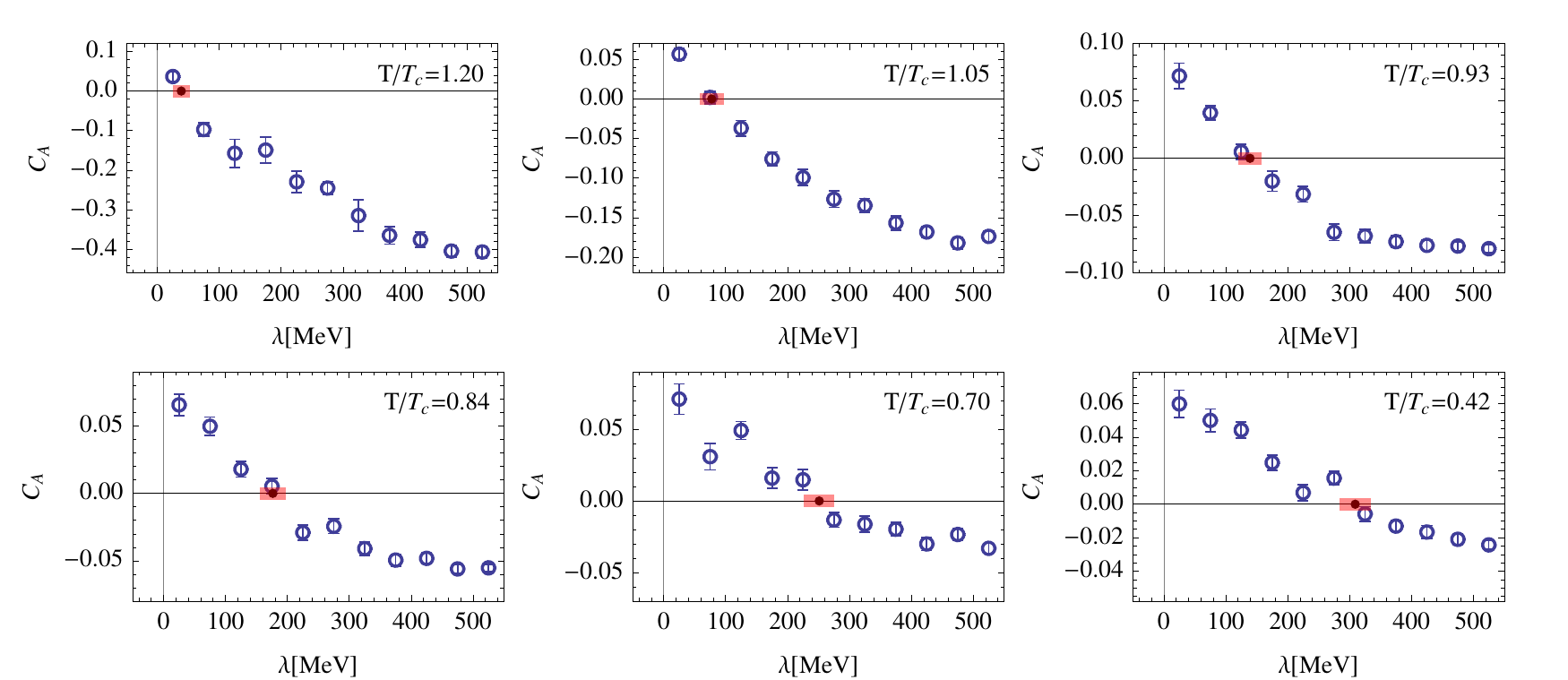}
\includegraphics[width=5.7cm]{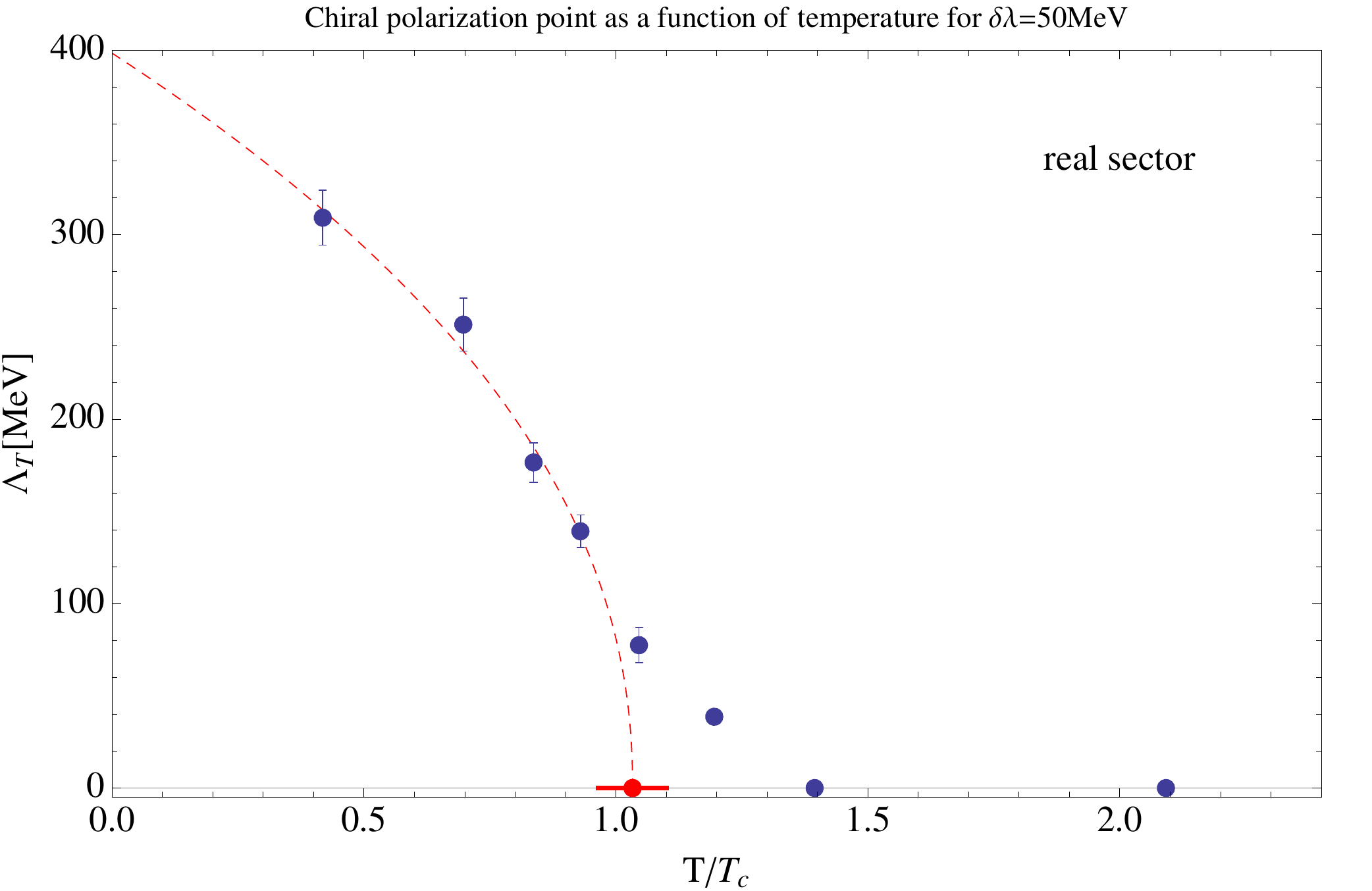}
}
\caption{Left panel: Correlation coefficient as a function of eigenmode scale.
Right panel: Chiral polarization scale as a function of temperature. The
dashed line indicate the fit discussed in the text.}
\label{fig:lam}
\end{figure}

To compute the chiral polarization scale, we bin the eigenvalues in bins of width
$\delta\lambda = 50\MeV$ and compute the average $C_A(\lambda)$ for each bin. In the left
panel of Fig.~\ref{fig:lam} we plot the average $C_A$ as a function of $\lambda$ for
each of ensemble where $C_A(\lambda=0) > 0$; these are the ensemble with $T/T_c\le 1.20$
where the low-lying modes are polarized. Note that, as in the zero temperature case,
at a sufficiently large $\lambda$ the modes become anti-polarized. The chiral
polarization scale is computed for each ensemble using a simple linear fit for the
bins that bracket the transition scale. The error bars are determined using the
jackknife method. The results are presented in the right panel of Fig.~\ref{fig:lam}.
As expected, the chiral polarization scale decreases as we increase the temperature.
A bit surprising is the fact that it does not vanish at $T=T_c$. Before we discuss
the cause for this phenomenon, we note that if we fit to a simple ansatz, 
$\Lambda(T) = \alpha (T-T_c)^\beta$, using $T_c$, $\alpha$ and $\beta$ as free parameters,
we find that the value of $T_c$ extracted from this fit is very close to the real
value. In this fit we only use the data points with $T<T_c$.

\begin{figure}[!t]
\center{
\includegraphics[width=12cm]{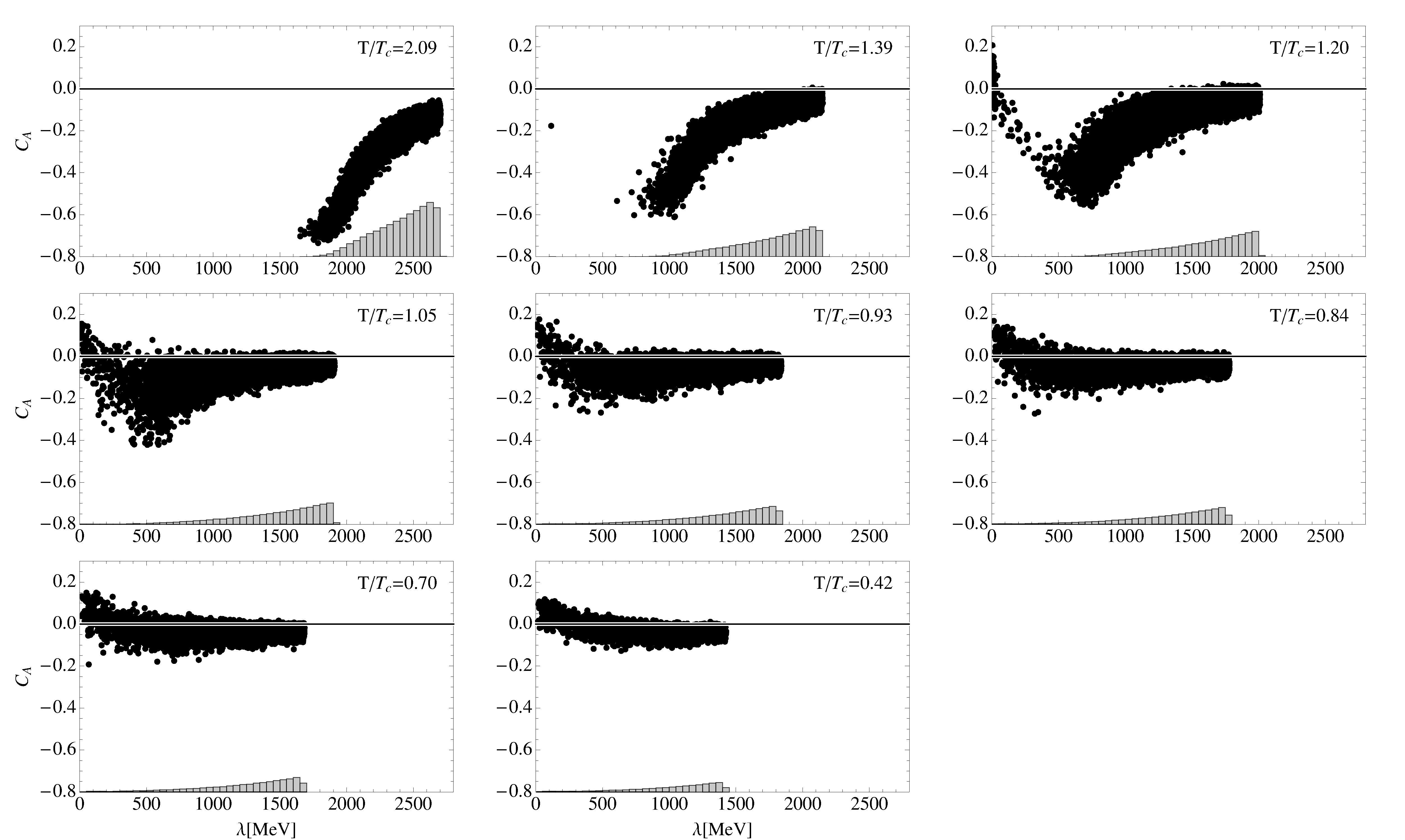}
}
\caption{Correlation coefficient versus eigenmode scale for all ensembles in this
study. The band below indicates the spectral density $\rho(\lambda)$. }
\label{fig:ca}
\end{figure}

The reason chiral polarization scale does not vanish as we cross the phase transition
is connected to the fact that the spectral density at the edge of the spectrum, 
$\rho(\lambda=0)$, does not vanish immediately above $T_c$. These near-zero modes are polarized and
lead to a non-vanishing chiral polarization scale. To show this, in Fig.~\ref{fig:ca} we
plot the eigenvalue and correlation coefficient for each mode in our ensembles.
Note that at low temperature $\rho(0)>0$ as required by Banks-Casher relation in
a phase where chiral symmetry is spontaneously broken; these near-zero modes are mostly 
polarized ($C_A>0$). The standard expectation is that above $T_c$ the Dirac spectrum 
develops a gap  and the symmetry is restored. This is indeed consistent with what we observe 
for $T>1.4 T_c$ and the modes at the low end of the spectrum are strongly anti-polarized. 
However, at intermediate temperatures, $T_c<T<1.4 T_c$, the situation is more complex: as
remarked earlier, a small density of near zero modes remains and these modes are polarized.
The existence of near-zero modes above deconfinement transition was observed earlier~\cite{Edwards:1999zm}.
We emphasize here that these are not zero modes, which are easy to
distinguish when using overlap operator. We removed the zero-modes from our analysis.
 

\section{Conclusions}
\label{sec:conclusions}

We computed the chiral transition scale as a function of temperature for pure gauge
configurations. We find that the scale decreases as we approach the phase transition. 
If we extrapolate from the confined phase, the polarization scale seems to vanish very
near $T_c$. However, direct calculations show that it vanishes for $T\approx 1.4 T_c$.
The discrepancy is due to the presence of near-zero modes at temperatures as high 
as~$1.2 T_c$. These modes are polarized causing the polarization scale to be non-zero. This
suggests that the deconfinment temperature and the chiral restoration 
scale---defined via the condition that $\rho(0)=0$---differ in the pure gauge theory.


The near zero modes, which are connected to chiral symmetry breaking, are almost all
(weakly) polarized. A more precise framework 
for discussing the polarization properties of the Dirac eigenmodes and their connection
to chiral symmetry breaking is presented in~\cite{Alexandru:2012sd}.

\medskip
\noindent{\bf Acknowledgments:} The computational resources for this project were provided in 
part by the George Washington University IMPACT initiative. 
This work is supported in part by the 
DOE grant DE-FG02-95ER-40907 and NSF CAREER grant PHY-1151648.

\bibliographystyle{JHEP}
\bibliography{my-references}

\providecommand{\href}[2]{#2}\begingroup\raggedright\begin{thebibliography}{10}

\bibitem{Banks:1979yr}
T.~Banks and A.~Casher, {\it {Chiral Symmetry Breaking in Confining Theories}},
   {\em Nucl.Phys.} {\bf B169} (1980) 103.

\bibitem{Schafer:1996wv}
T.~Schafer and E.~V. Shuryak, {\it {Instantons in QCD}},  {\em Rev.Mod.Phys.}
  {\bf 70} (1998) 323--426, [\href{http://xxx.lanl.gov/abs/hep-ph/9610451}{{\tt
  hep-ph/9610451}}].

\bibitem{Horvath:2001ir}
I.~Horv\'ath, N.~Isgur, J.~McCune, and H.~B. Thacker, {\it {Evidence against
  instanton dominance of topological charge fluctuations in QCD}},  {\em Phys.
  Rev.} {\bf D65} (2002) 014502,
  [\href{http://xxx.lanl.gov/abs/hep-lat/0102003}{{\tt hep-lat/0102003}}].

\bibitem{Draper:2004id}
T.~Draper, A.~Alexandru, Y.~Chen, S.-J. Dong, I.~Horv\'ath, {\em et.~al.}, {\it
  {Improved measure of local chirality}},  {\em Nucl.Phys.Proc.Suppl.} {\bf
  140} (2005) 623--625, [\href{http://xxx.lanl.gov/abs/hep-lat/0408006}{{\tt
  hep-lat/0408006}}].

\bibitem{Alexandru:2010sv}
A.~Alexandru, T.~Draper, I.~Horv\'ath, and T.~Streuer, {\it {The Analysis of
  Space-Time Structure in QCD Vacuum II: Dynamics of Polarization and Absolute
  X-Distribution}},  {\em Annals of Physics} {\bf 326} (2011) 1941--1971,
  [\href{http://xxx.lanl.gov/abs/1009.4451}{{\tt arXiv:1009.4451}}].

\bibitem{Alexandru:2010rg}
A.~Alexandru, T.~Draper, I.~Horv\'ath, and T.~Streuer, {\it {Absolute Measure
  of Local Chirality and the Chiral Polarization Scale of the QCD Vacuum}},
  {\em PoS} {\bf LATTICE2010} (2010) 082,
  [\href{http://xxx.lanl.gov/abs/1010.5474}{{\tt arXiv:1010.5474}}].

\bibitem{Alexandru:2011yy}
A.~Alexandru and I.~Horv\'ath, {\it {How Self-Dual is QCD?}},  {\em Phys.Lett.}
  {\bf B706} (2012) 436--441, [\href{http://xxx.lanl.gov/abs/1110.2762}{{\tt
  arXiv:1110.2762}}].

\bibitem{Alexandru:2011tu}
A.~Alexandru and I.~Horv\'ath, {\it {Absolute X-distribution and
  self-duality}},  {\em PoS} {\bf LATTICE2011} (2011) 268,
  [\href{http://xxx.lanl.gov/abs/1111.3897}{{\tt arXiv:1111.3897}}].

\bibitem{Guagnelli:1998ud}
{\bf ALPHA} Collaboration, M.~Guagnelli, R.~Sommer, and H.~Wittig, {\it
  {Precision computation of a low-energy reference scale in quenched lattice
  QCD}},  {\em Nucl. Phys.} {\bf B535} (1998) 389--402,
  [\href{http://xxx.lanl.gov/abs/hep-lat/9806005}{{\tt hep-lat/9806005}}].

\bibitem{Karsch:2003jg}
F.~Karsch and E.~Laermann, {\it {Thermodynamics and in medium hadron properties
  from lattice QCD}},  \href{http://xxx.lanl.gov/abs/hep-lat/0305025}{{\tt
  hep-lat/0305025}}. Prepared for Quark-Gluon Plasma III, R. Hwa (ed.).

\bibitem{Chandrasekharan:1995gt}
S.~Chandrasekharan and N.~H. Christ, {\it {Dirac spectrum, axial anomaly and
  the QCD chiral phase transition}},  {\em Nucl.Phys.Proc.Suppl.} {\bf 47}
  (1996) 527--534, [\href{http://xxx.lanl.gov/abs/hep-lat/9509095}{{\tt
  hep-lat/9509095}}].

\bibitem{Edwards:1999zm}
R.~G. Edwards, U.~M. Heller, J.~E. Kiskis, and R.~Narayanan, {\it {Chiral
  condensate in the deconfined phase of quenched gauge theories}},  {\em
  Phys.Rev.} {\bf D61} (2000) 074504,
  [\href{http://xxx.lanl.gov/abs/hep-lat/9910041}{{\tt hep-lat/9910041}}].

\bibitem{Alexandru:2012sd}
A.~Alexandru and I.~Horv\'ath, {\it {Spontaneous Chiral Symmetry Breaking as
  Condensation of Dynamical Chirality}},
  \href{http://xxx.lanl.gov/abs/1210.7849}{{\tt arXiv:1210.7849}}.

\end{thebibliography}\endgroup

\end{document}